# Security and Privacy Considerations for Machine Learning Models Deployed in the Government and Public Sector

Nader Sehatbakhsh, Ellie Daw, Onur Savas, Amin Hassanzadeh, Ian McCulloh

Accenture, Washington DC, USA
{nader.sehatbakhsh, ellie.daw, onur.savas, amin.hassanzadeh, ian.mcculloh}@accenture.com

**Abstract**

As machine learning becomes a more mainstream technology, the objective for governments and public sectors is to harness the power of machine learning to advance their mission by revolutionizing public services. Motivational government use cases require special considerations for implementation given the significance of the services they provide. Not only will these applications be deployed in a potentially hostile environment that necessitates protective mechanisms, but they are also subject to government transparency and accountability initiatives which further complicates such protections.

In this paper, we describe how the inevitable interactions between a user of unknown trustworthiness and the machine learning models, deployed in governments and public sectors, can jeopardize the system in two major ways: by compromising the integrity or by violating the privacy. We then briefly overview the possible attacks and defense scenarios, and finally propose recommendations and guidelines that once considered can enhance the security and privacy of the provided services.

## Introduction

Machine Learning (ML) technologies are beginning to permeate all industries, both in the public and the private sectors. AI and ML enable more complex computations to be deployed to a wider audience in a variety of useful manners including financial markets, transportation, national security, healthcare, and many other applications.

Due to these potentials in ML, there are many opportunities that could benefit governments and public sectors. However, to effectively utilize these systems, there are a number of critical considerations that must be taken into account including but not limited to ethics, fairness, performance, and trustworthiness. Although all of these are important to address, *security and privacy* are paramount because governments typically have access to sensitive citizen data and provide services that are critical to quality-of-life. Thus, the important question before deploying any ML model in a security critical environment is whether the deployed model is safe against potentially abusive or malicious users who interact with the system?

To address the problem of using ML models in critical systems, in this paper we first provide three motivational use cases from healthcare, area protection, and census. Building upon these use cases, we then provide an overview of the existing threats on the deployed ML models in an untrusted and potentially hostile environment. Many government and public sector technologies compliance standards, for example FedRAMP in the United States, to leverage best-practice security considerations and mitigate common vulnerabilities. Additionally, an ML service will have users who are able to interact with it to utilize the outcome of a model; for example, a citizen submits information to a service to see if he qualifies for benefits. Next, we will outline how *the way a user interacts with the system* leads to potential security and privacy issues even if the deployed ML product or service conforms to standards requirements. We emphasize that these threats are not only limited to the model, but they can also target the data and hence violate user privacy. We also taxonomize different classes of attacks and elaborate on the existing countermeasures to mitigate these attacks. We conclude with recommendations and guidelines on how ML systems should be securely deployed and evaluated, and what proactive measures the policy makers can take.

## Motivation

There exist use cases that if not carefully designed and developed, the intrinsic properties of machine learning systems will be exploited by honest-but-curious or adversarial entities resulting in fraud, waste, and abuse (Accenture, 2019). Consider government agencies that provide citizen services such as subsidized health coverage, affordable housing, and student or farmer loans. In many cases, the government is the facilitator and the delivery of the service is done via a commercial entity such as a financial institution or an insurance company. Agents and brokers are also frequently used as re-sellers of these services. Machine learning can help to quantify the risk and sustain the delivery of these services that can be beneficial to all parties involved. However, ML models, whose intention is to aid in decision making, can be stolen. For example, a malicious broker can learn both how the model computes risk and the parameters of the model, e.g., by repeatedly querying the model despite having an explicit need. By querying, he can learn the risk thresholds below which the loan will not be forwarded to the collection agency and then use this knowledge to his advantage by offering lower rates.

Consider another example where a government agency protects wildlife or controls a hazardous area by prohibiting people from entering. Traditionally, guards or law enforcement have to patrol the perimeter, watch security footage, or rely on simple triggers that are indicative of activity.

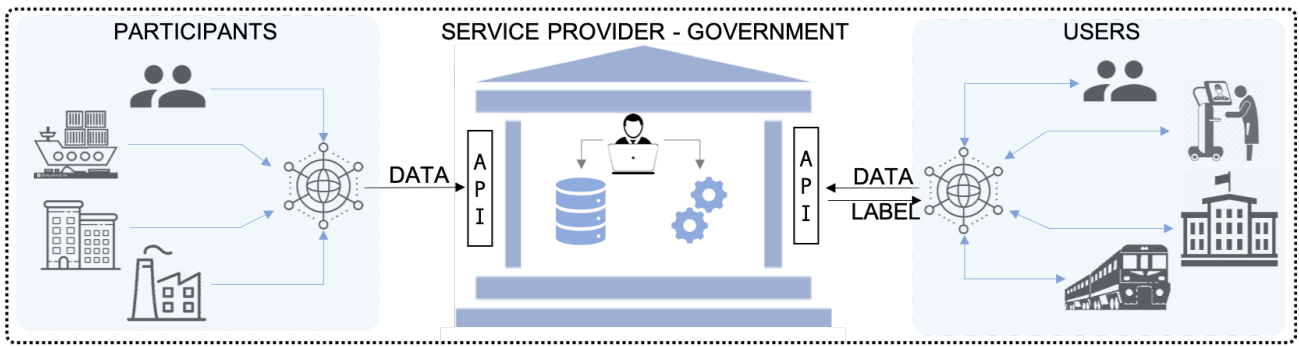

*Figure 1. A typical machine learning ecosystem including four major roles: participants, service provider, developer, and users.*

However, this process is labor-intensive, prone to error, and cannot distinguish between animals and people. With recent advances in deep neural networks that can achieve above-human detection rates, it has become easier to detect activity and identify human presence in videos. Hence, the government agency can deploy an ML model to decrease labor intensive patrols or hours of tedious video examination, thus automating detection of human activity at a fraction of the cost. However, an adversarial actor with knowledge of the deployed neural network can evade detection by simply providing an "adversarial input." For example, in (Thys, 2019), a person holding a specific painting can walk by the camera without being detected creating a security breach.

One example of a model with wide citizen impact is within government agencies that are responsible for collecting census and economic data to better understand the communities they serve. In addition to releasing summary statistics, these agencies can deploy predictive models to aid policy makers, NGOs, commercial sector, and the citizens in making informed decisions. Consider a model that predicts areas of housing growth (or decline). In a small community, real-estate agents could correlate the outputs of such models with relevant datasets to potentially identify specific houses or small neighborhoods. In a large city like New York City, such a privacy compromise may be difficult, but in a small city it could give an unfair advantage to a dishonest agent.

## System and Attack Models

*Machine learning* (ML) is the science of teaching machines how to act and improve their learning over time without being programmed. An ML system automates the analysis of (typically) large datasets, producing models or decision procedures reflecting general patterns found in that data. The learning process can be either continuous where training the system is incremental and data instances are collected sequentially, or offline where a batch of training data is available to train the system before deployment. ML techniques are commonly divided into three classes, characterized by the nature of the data available for analysis and the type of supervision during training. These classes are *supervised learning*, *unsupervised learning,* and *reinforcement learning*. Depending on the task and its complexity, ML models can range from simple linear regression models to deep neural networks. Regardless of the ML model and the learning technique, as shown in Figure 1, there are **four** major roles in an ML system (Mitchell, 1997):

- *Participants* are one or a group of public or private entities that provide the training data. In Figure 1, participants are shown as individuals such as private citizens and patients, or a public sector.
- *Model/Service Provider* is a platform owned by an entity that gathers the training dataset to train the model. Once the model is trained, it also holds the *parameters (*i.e., the weights and the structure of the ML model). Further, the provider also acts as a server, providing an interface for the model to be used for prediction. Note that, depending on the use-case, the server and participants may or may not be a same entity.
- *Developer/AI Expert* is the entity that interacts with the server to train the model by providing the training algorithms and hyperparameters. Typically, the service provider and the developer are the same. However, the developer can be an external entity (e.g., a contractor). In Figure 1, the developer is shown as a part of the service provider entity.
- *User* is a person or an entity such as hospitals, transportation or education system, that interacts with the trained model through some interface in order to process their data using the ML model.

**Attack Model.** At a high-level, attacks on machine learning systems can be launched in two different phases (Papernot, 2018): *training*- where the model is being trained; and *inference*- where the trained model is applied to a task (e.g., labeling an image, detecting a pattern, etc.). While there are several interesting challenges for securing the system during the training (Yang 2019), the focus of this paper is on the attacks that target the model during the *inference*, i.e., when the model is deployed and being used by the, potentially untrusted, users.

Attacks on an ML model during i*nference* can be further categorized into two main types (Papernot 2018): ***integrity attacks***– where the goal is to manipulate the functionality of the model, i.e., to change the output label or outcome of the model; and ***privacy/confidentiality attacks***- where the goal is to *understand* the data used in the model (i.e., either the parameters in the model or the data that was used during the training of the model). These attacks are shown in Figure 2.

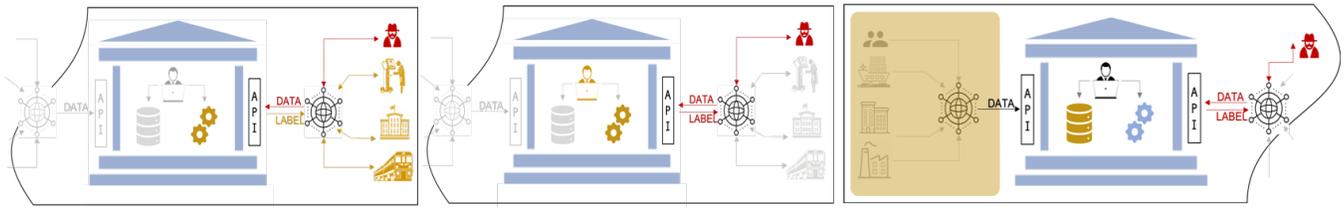

*Figure 2. Model evasion attack (left), model extraction attack (middle), membership inference attack (right). Red represents the malicious user, while yellow depicts the impacted entities (victims).*

In both attack models, the assumption is that the **user is the adversary**. This adversary can interact with the model through an application program interface *(API)* by sending input and receiving the label(s). It is important to mention that since the model is already trained, the adversary (user) *cannot impact the parameters* of the model, thus attack scenarios such as poisoning, backdoor, etc., which poison the training dataset, either during the training or continuously, (Biggio 2012) are not possible. As we will later see, depending on the attack, the level of knowledge that the adversary has about the model or its training data can range between fully unknown (*black-box*) to fully-known (*white-box*), allowing the adversary to launch different types of attacks.

It is worth mentioning that, to further simplify the discussion, we provide examples for attacking *supervised learning* systems due to their popularity and availability. However, similar attacks can be adopted for *unsupervised* and/or *reinforcement* learning systems. Lastly, we assume that in addition to providing the final class label, the API may also provide *confidence values* for all the possible class labels.

**Integrity Attacks.** The goal in these attacks is to affect the outcome of the system. Since in ML systems the outcome is typically a label, the integrity attacks (during inference) aim to engineer the input in such a way that, ultimately, causes a *misclassification.* Integrity attacks during the inference phase are called *evasion* attacks (Šrndic 2014).

Such attacks can be categorized into two approaches (Bae 2018): **white-box** where the adversary has semi- or full-knowledge about the model and its parameters, or **black-box** where the adversary knows nothing about the system and can only interact (i.e., sending inputs and receiving labels with corresponding confidence values). While it is often difficult to obtain, white-box access is not unrealistic. For instance, distributed ML models trained on data centers are compressed and deployed to smartphones, where an adversary may be enabled to reverse engineer the model's internals (e.g., its parameter values) and thus obtain white-box access (Papernot 2018).

Regardless of the access level, evasion attacks can be launched by either *direct* manipulation of the input data (called *norm-based attacks*) or, alternatively, by indirect manipulation of the data called *physical attacks* (Papernot 2018). For example, wearing glasses to fool a face recognition system is a physical evasion attack, and adding white gaussian noise to an image is an example of direct manipulation. Furthermore, depending on the outcome, the evasion attacks can be either *targeted* in which the adversarial input is designed to generate a specific outcome or *untargeted* where the objective is misclassification (Bae 2018).

The main idea behind an evasion attack is to create input samples, also called *adversarial examples*, that once processed by the model create an incorrect label. This is usually done by manipulating input features near the model's decision boundaries which, in turn, fools the classifier. To create such samples, typically standard non-linear optimization techniques are used (Goodfellow 2014).

**Privacy Attacks.** The goal of this attack is to compromise the privacy of the ML model by either: a) stealing the model which is called **model extraction** (Tramèr 2016); or b) retrieving the data used during training.

Retrieving training data can be categorized as: 1) **membership inference** where the adversary finds out whether or not a specific record was present during the training (Shokri 2017), and 2) **model inversion** where the adversary finds a specific piece of information in the training dataset using the information that he already has such as parts of the dataset, model, labels confidence values, etc. (Fredrikson 2015).

Another important use case of privacy attacks is that *they can be used as an intermediate step to launch a (white-box) integrity attack.* For example, to launch an evasion attack the adversary can first perform model extraction, hence achieving white-box access to the model, and then continue on to the evasion attack (Fredrikson 2015).

Like evasion, attacks on privacy can be either white- or black-box depending on the adversary's knowledge. In addition to accessing the model, knowledge about the training dataset can also be critical in a successful attack, especially for membership inference and model inversion. Recent work has shown that successful privacy attacks can be launched even with no knowledge about the model or its training data, further confirming the power of these attacks (Salem 2018).

The main idea behind model extraction is solving a system of equations to find the unknowns (i.e., parameters in the model) when the final output value is some linear or non-linear combination of the weights and the inputs. To solve this system, an adversary repeatedly queries the ML model (i.e., sends many inputs), and receives the label and the confidence values. The adversary then uses these values to solve the equations (i.e., where the inputs and the outputs are known, and weights are unknown). Interestingly, finding weights in this system of equations is almost analogous to finding the weights during the training, techniques used in training, such as stochastic gradient decedent, can be used by the adversary to find the weights (Tramèr 2016).

Membership inference and model inversion attacks use an algorithm that takes the input data and the confidence values and generates either the actual training record (fully or partially) for model inversion, or a binary value indicating whether or not the record was part of the training. Then, the adversary can use the outputs to *train* his attack machine (Shokri 2017). Here, the ML attack machine is an *inverted* version of the actual ML model.

## Attack Countermeasures

Integrity attacks generally arise from a lack of model robustness, which allows a user to manipulate the outcome of an ML system (Nicholas 2017). These types of attacks can be difficult to protect against, but the first step should be to evaluate a system against adversarial examples. Some ecosystems may be less sensitive to an occasional misclassification, but in other high-stakes environments it is important that adversarial avenues are explored before deployment. After examples are identified, a model can be trained to ignore them, called *adversarial training* (Tramèr 2018).

Privacy attacks arise because most systems allow a user to make multiple queries, and a malicious user can learn from collecting numerous query responses. One possible countermeasure is to monitor the model querying API for possible misuse. Another important issue that enables privacy attacks is *overfitting* (Shokri 2017). Practical countermeasures for this issue rely on using standard training techniques to avoid overfitting and/or using ensemble learning to avoid leakage (Salem 2018).

Privacy attacks against participant data can also arise from a lack of anonymization and may become easier as the size of the training dataset decreases. In order to prevent these attacks, it is important to take measures to keep training data private. This can be done in several ways:

- *Differential privacy* aggregates data in a way that individual privacy cannot be compromised. Typically, this is done by adding noise to a dataset.
- *Homomorphic encryption* is a technique that allows computations to be done on encrypted data without revealing the unencrypted data.
- *Multi-party computation* is a technique that enables multiple parties to jointly perform an operation without any of them having access to the full dataset.

Specific details for these techniques can be found in the literature. Any of these techniques can be employed to ensure privacy of the participants during inference.

## Discussions and Recommendations

The technical countermeasures discussed above are most beneficial if they are considered during the design phase of a project. Thus, governments and public sectors should ensure that there is sufficient time and space to accommodate such careful considerations from the beginning.

After considering the relevancy of the technical countermeasures to improve the robustness of public sector ML-based services, we recommend further measures be adopted into the ecosystem to ensure that security and privacy implications are continuously evaluated. Although there aren't many current instances where ML systems are implemented in government services, the motivating use-cases discussed above provide great context for the gravity of impact such systems hold. Various motivating use-cases relate to critical services and can contribute to the quality of life for citizens. Thus, it is paramount that use-cases deploy systems that are robust against hostile users, but also put in place an environment where concerns can be revisited at any time.

In order to better understand and prepare for specific use-cases, stakeholder requirements, and respective security and privacy impacts, governments should facilitate the formation of communities of interest for security and privacy of machine learning. These communities would consist of all stakeholders and would aim to serve the community by collecting diverse and comprehensive knowledge, then evaluating for security and privacy concerns. Similar to other cybersecurity-related groups, such a community could recommend best-practices and advise policy-makers during the process of forming laws and regulations, where relevant, in the interest of citizens. For example, privacy interest groups have been involved in shaping the creation of privacy laws and regulations around the world, such as General Data Protection Regulation (GDPR).